\documentclass{aa}  

\usepackage{graphicx}
\usepackage{txfonts}
\begin{document} 
	\title{Properties of Low Luminosity Afterglow Gamma-ray Bursts}
	
 \author{H. Dereli
          \inst{1}\thanks{Eramus Mundus Joint Doctorate IRAP PhD student}
          \and
          M. Bo\"er 
          \inst{1}\thanks{Corresponding author. Michel.Boer@unice.fr}
          \and
          B. Gendre
          \inst{1}$^{, }$\inst{2}$^{, }$\inst{3}
          \and
          L. Amati
          \inst{4}
          \and 
          S. Dichiara
          \inst{5}
          }
 \institute{ARTEMIS (CNRS UMR 7250, OCA, UNS); Boulevard de l'Observatoire, CS 34229, 06304 Nice Cedex 4, France
         \and
         			Etelman Observatory, Bonne Resolution, 00802 St Thomas, V.I., USA
         \and
         			University of  the Virgin Islands, 2 John Brewer's Bay road, 00802, St Thomas, V.I., USA
         \and
              IASF-Bologna/INAF; Area della Ricerca di Bologna, Via Gobetti 101, 40129 - Bologna, Italy
         \and
             University of Ferrara; Via Savonarola 9, 44121 Ferrara, Italy  
             }
   \date{Received ---; Accepted ---}

 
 \abstract
   {}
   {We characterize a sample of Gamma-Ray Bursts with low luminosity X-ray afterglows (LLA GRBs), and study their properties.}
   {We select a sample consisting of the 12\% faintest X-ray afterglows from the total population of long GRBs (lGRBs) with known redshift. We study their intrinsic properties (spectral index, decay index, distance, luminosity, isotropic radiated energy and peak energy) to assess whether they belong to the same population than the brighter afterglow events.}
   {We present strong evidences that these events belong to a population of nearby events, different from that of the general population of lGRBs. These events are faint during their prompt phase, and include the few possible outliers of the Amati relation. Out of 14 GRB-SN associations, 9 are in LLA GRB sample, prompting for caution when using SN templates in observational and theoretical models for the general lGRBs population.}
   {}


   \maketitle
%

\section{Introduction}
\label{sec_intro}

Gamma-ray bursts (GRBs) are the most luminous events in the Universe, with isotropic luminosity between $10^{49}-10^{52} ${erg.s}$^{-1}$ \citep{mes06}. 
GRBs display two components: the prompt emission, followed by an afterglow \citep{ree92, mes97, pan98}, both observed at all wavelengths \citep{cos97, van97, fra97}. In X-rays, the afterglow light curve can be described as a steep-flat-steep broken power law \citep{nou06}. The first part (steep decay) has been associated with the prompt phase \citep{wil07, zha06} while the central engine is still active; the rest of the afterglow is due to the dynamics of the interaction of the jet with the surrounding medium. 

Several studies have been made on GRB samples \citep[e.g.][]{mel14}, but in general they do not address specific properties. Usually the authors focus on complete samples in order to derive broad properties. 

These properties are then used to define the unknown physical properties of an archetypal GRB. In this work, we consider that the population of long GRBs (hereafter lGRBs) may hide various sub-types of GRBs;  thus it is important to check for the existence  of different populations in the sample, and, should this happen, how the previous conclusions apply to the whole populations. This has already been shown with the class of ultra-long GRBs \citep{gen13,zha14, boe14}. 

In the past, several GRBs featuring faint prompt emission have been observed: GRB 980425 \citep[e.g.]{gal98, kul98, pia06}, GRB 031203 \citep[e.g.][]{mal04, sod04, watson04}, GRB 060218 \citep[e.g.]{cam06, maz06, sod06, virgili08} and GRB 100316D \citep[e.g.]{fan11, sta11}. Theoretical work has been done on the standard model in order to explain these events \citep[e.g.]{dai07, bar15}, and from their obvious properties (a low luminosity) several authors have pointed out that only very few faint events are detected \citep[e.g.][]{imerito}. All these studies are based on the properties of single events (despite the fact that GRB 060218 and GRB 980425 have very different properties, for instance), and so far the global sample of {\em Swift} prompt faint events has not been studied. Moreover, to our knowledge, despite a different approach and properties, no systematic study (and actually no individual) has been made by selecting a sample on the basis of the faintness of the afterglow. 

This is the purpose of this work, and in the following we call GRB displaying dim afterglow (according to our criteria, Low Luminosity Afterglow GRBs (LLA GRBs). We explicit the criteria to build a consistent sample of LLA GRBs and we derive its properties. Then, we use a control sample based on different bursts to check whether they form a class different from that of normal lGRBs.

This paper is organized as follows: In section \ref{sec_sample}, we present the LLA GRB sample and we describe how we selected it. In section \ref{sec_properties}, we discus the possible biases and basic properties of our sample. In section \ref{sec_discu}, we discuss our results, before concluding in section \ref{sec_conc}. In the following, all errors are quoted at the 90\% confidence level, and we used a standard flat $\Lambda$CDM model with $\Omega_m$=0.3 and $H_0 = 72 km s^{-1} Mpc^{-1}$. We also used the standard notation $F \propto t^{-\alpha} \nu^{-\beta}$.

\section{Definition of the sample}
\label{sec_sample}

We took into account all bursts with a measured redshift observed before 2013, February the 15$^{th}$, without consideration of the detector triggered by the event and/or observing it. We have used the list compiled by Greiner\footnote{http://www.mpe.mpg.de/$\sim$jcg/grbgen.html}.  
This leads to a first sample of 283 sources which have been observed at X-ray wavelengths, including short and long GRBs. As we are interested only in the later, we have to exclude sGRBs: to that purpose we used the method described in \citet[][; this method classifies short GRBs all burst with a duration less than 2 seconds in the rest frame, with additional criteria on the afterglow]{sie14} to reject them, leaving 254 long bursts in the global sample.

As the analysis of the bursts that happened prior to 2006 was already performed by \citet{gen08}, we describe here the method followed for the {\em Swift} bursts only. We retrieved the XRT light curve from the online {\em Swift} repository\footnote{http://www.swift.ac.uk/xrt$\_$curves} \citep{eva09}. 

Comparing flux light curves is a complex task, and need a careful estimation of the spectral index and the count-to-flux conversion factor. The estimation of these two parameters are done automatically for the online repository light curve, using standard models that may fail for various reasons, or not correspond to our needs (for instance, a spectral index estimated with some data taken before the end of the plateau phase, see below). We thus cannot use directly the data downloaded from the online repository, and needed to estimate independently the spectral index and the count-to-flux conversion factor. 

For this purpose, we also downloaded the raw data from the archives, and applied to them the calibration released in May 2013 (using the Swift software, distributed as part of the HEASOFT package, version 6.12). We then extracted a spectrum using the task {\em xselect}, also part of the HEASOFT package\footnote{http://heasarc.gsfc.nasa.gov/heasoft/}, and fit it with a power law model absorbed twice, in the host galaxy and in the Milky-Way. The N$_H$ value of the Galaxy was set to the value given by the Leiden/Argentine/Bonn (LAB) Survey of Galactic HI \citep{kal05}, while the one for the host was let free to vary at the host redshift.  Finally we compared this best fit model with the one from the automated analysis pipeline: in case of inconsistency we recomputed the flux light curve using the conversion found from our best fit model.

\begin{figure*}
\begin{center}
\includegraphics[width=16cm]{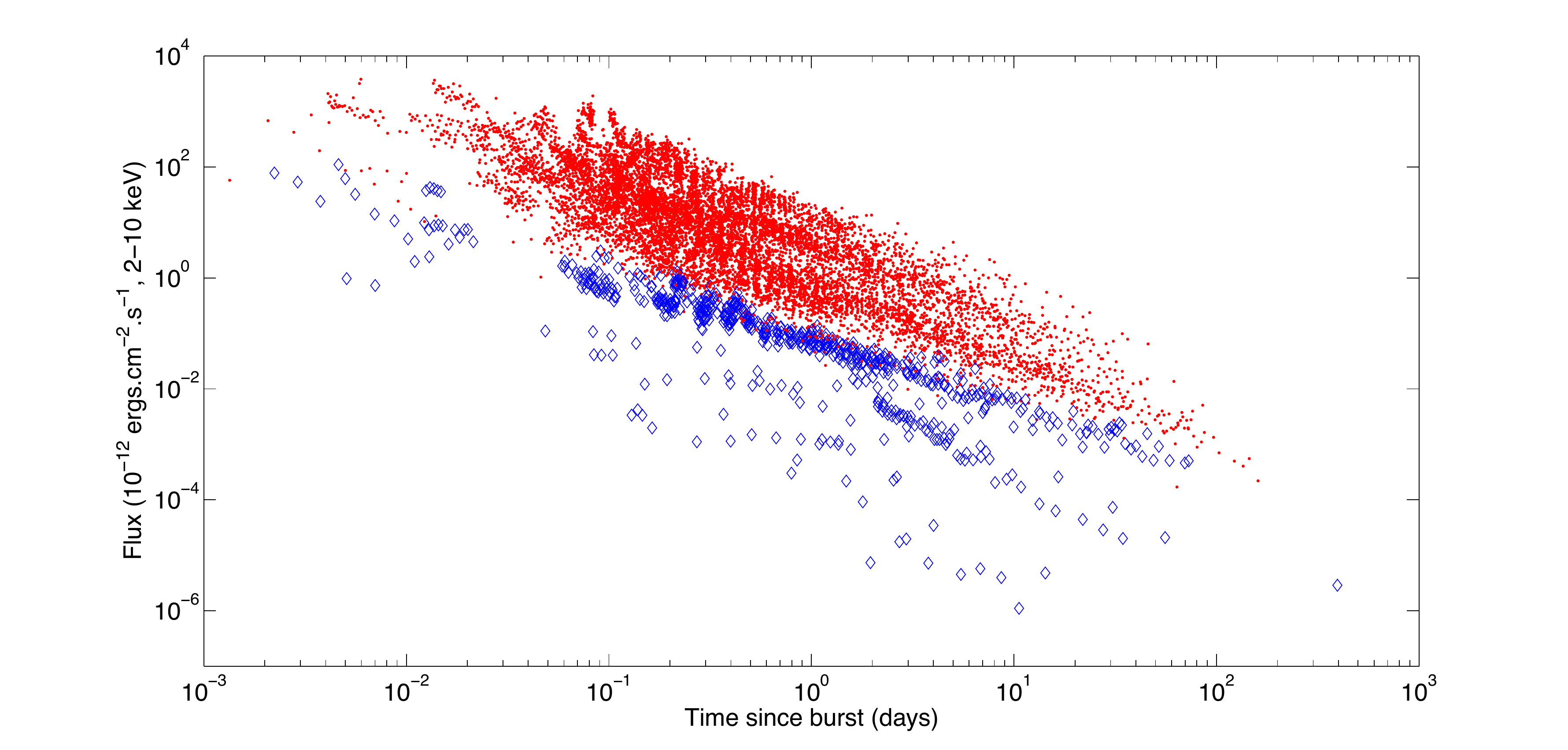}
\caption{The light curves of all sources, corrected for distance effects (see text) and rescaled at a common redshift z=1. LLA events are shown with blue diamonds, the control sample is shown with red dots.\label{fig_bg}}
\end{center}
\end{figure*}

Once the flux calibration has been checked and eventually corrected, we selected the afterglow part of the light-curve. We followed the method of \citet{gen08}, removing from the light curves all emission present before the end of the plateau phase and all flaring emissions. This net light curve was then corrected taking into account the cosmological effects including the K-correction.

We worked on a "flux" light curve, rescaling all light curves to a common distance of $z = 1$. As stated in \citet{gen05}, this allows a smaller uncertainty on the final light curves. One may wonder, now with the precise cosmology parameters measured by Planck  \citep{hin13, ade13} whether this is really needed; the reason is that the uncertainty is introduced by the K-correction (and not by the distance correction), which is very sensitive to the spectral index:

\begin{equation}
K \propto (1+z)^\beta
\end{equation}

As an example, with a redshift of 4 and a precision of $1.0 \pm 0.3$ for $\beta$, the uncertainty on K is $5 \pm 3$, i.e. 60\%. Rescaling to z $=1$ leads to $2.5 \pm 0.7$, i.e. an uncertainty of 28\% : this method reduces the scattering induced by the uncertainties on the measurements, allowing for a more precise selection of the sample.

Being interested on LLA events, we defined two template afterglows with a decay index of 1.2 and 1.4 respectively (corresponding to the typical values expected with $p \sim 2.3$ where p is the power law index of the accelerated electrons in the cases of wind and interstellar media); we set \textit{a priori} the limit at $F = 1 \times 10^{-13}$ erg cm$^{-2}$ s$^{-1}$ one day after the burst. 

There are two reasons for this choice: first, we are interested in the low luminosity part, and thus we chose a flux significantly lower than the mean observed flux for the afterglows present in Fig. \ref{fig_bg}; second, as noted in \citet{gen08}, there are bursts with a low luminosity that seems to not follow the properties of the other groups. These events represent about 10\% of the total burst population, which turns to a limiting flux of about $10^{-13}$ erg cm$^{-2}$ s$^{-1}$ one day after the burst. Additionally, the flux cut-off $10^{-13} \text{erg cm}^2\text{ s}^{-1}$ at 1 day corresponds roughly to the lowest afterglow luminosity at one day of the unbiased sample of \citet{dav12a} (there in as seen from Figure 2).

All bursts with an afterglow light curve entirely below these two templates were part of LLA GRBs sample; the others are used as a control sample. The result of the selection is displayed in Fig. \ref{fig_bg}. 

The final sample includes 31 events that are listed in Table \ref{table_sample}, representing about 12\% of all lGRBs considered here. Table \ref{table_sample} displays the GRB name, redshift, galactic and host N$_H$, galactic and host A$_V$, the afterglow temporal and spectral indexes, the isotropic and peak energies, and the T$_{90}$ duration (the time during which 90\% of the energy of the prompt is emitted). For those afterglows displaying a break after the plateau phase (GRB 060614 and GRB 120729A), the decay index is indicated pre-break.

\begin{table*}
\centering
  \caption{The burst sample and its main characteristics (see text). The spectral and temporal indexes for GRBs before August 2006 are taken from \citet{gen08}. \label{table_sample}}
  {\scriptsize
  \begin{tabular}{lllllllllllll}
  \hline
GRB        & z      & \multicolumn{2}{c}{N$_H$} & \multicolumn{2}{c}{A$_V$} & \multicolumn{2}{c}{Afterglow} & logT$_a$  & E$_{\rm iso}$ &  E$_{\rm p,i}$ 	&T$_{\rm 90}$ &Ref.\\
           &        & Gal & Host                &  Gal & Host            &  Temporal & Spectral             & (s)    &(10$^{52}$erg) & (keV)	  &	(s) & \\
           &        & \multicolumn{2}{c}{(10$^{21}$cm$^{-2}$)} & \multicolumn{2}{c}{(mag)} & index  &  index &       &               &          &     & \\
\hline
GRB 980425 & 0.0085 & 0.428 & $\cdots$ &0.071 &1.73 & 0.10$\pm$0.06 & (0.8) &$\cdots$ & (1.3$\pm$0.2)$\times10^{-4}$ &55$\pm$21 & 18 & (1), (13) \\ 
GRB 011121& 0.36 & 0.951 & $\cdots$ & 0.061 & 0.38 & 1.3$\pm$0.03 & (0.8) & $\cdots$ & 7.97$\pm$2.2 & 1060$\pm$275 & 28 & (1), (14), (15) \\
GRB 031203& 0.105	 & 6.21 & $\cdots$ & 0.117 & 0.03 & 0.5$\pm$0.1& 0.8$\pm$0.1 &$\cdots$ & (8.2$\pm$3.5)$\times10^{-3}$ & 158$\pm$51 &40 &(1), (14) \\
GRB 050126 & 1.29 & 0.551 & (0.0) & 0.182 & $\cdots$ &1.1$^{+0.6}_{-0.5}$ & 0.7$\pm$0.7 & $\cdots$ & [0.4 - 3.5] & $>$201 & 24.8 & (17) \\
GRB 050223 & 0.5915 & 0.729	 & (0.0) & 0.078 & $>$2 & 0.91$\pm$0.03 & 1.4$\pm$0.7 &$\cdots$ & (8.8$\pm$4.4)$\times10^{-3}$	 & 110$\pm$55 & 22.5 &(2), (18) \\ 
GRB 050525 & 0.606 & 0.907 & 0.38$^{+9.1}_{-0.38}$ & 0.221 & 0.36$\pm$0.05 & 1.4$\pm$0.1 & 1.1$\pm$0.4 & 3.8 & 2.3$\pm$0.5 & 129$\pm$12.9 & 8.8 & (5), (19) \\
GRB 050801 & 1.38 & 0.698 & (0.0) & 0.989 & 0.3$\pm$0.18 & 1.25$\pm$0.13 & 1.84$^{+0.56}_{-0.53}$ & 3.2 & [0.27 - 0.74] & $<$145 & 19.4 & (5), (17) \\
GRB 050826 & 0.297 & 2.17 & 8$^{+6}_{-4}$ & 2.398 & $\cdots$ & 1.13$\pm$0.04 & 1.1$\pm$0.4 & 4.04 & [0.023 - 0.249] & $>$37 & 35.5 & (17) \\
GRB 051006 & 1.059 & 0.925 & (0.0) & 2.345 & $\cdots$ & 1.69$\pm$0.13 & 1.5$^{+0.44}_{-0.46}$ & 2.77 & [0.9 - 4.3] & $>$193 & 34.8 & (17) \\
GRB 051109B & 0.08 & 1.3 & $<$2 & 0.3 & $\cdots$ & 1.1$\pm$0.3 & 0.7 $\pm$0.4 & 3.14 & $\cdots$ & $\cdots$ & 14.3 & \\
GRB 051117B & 0.481 & 0.46 & (0.0) & 0.321 & $<1.4$ & 1.03$\pm$0.5 & (0.8) & $\cdots$ & [0.034 - 0.044] & $<$136 & 9.0 & (11), (17) \\
GRB 060218 & 0.0331 & 1.14 & 6$\pm$2 & 0.437 & 0.5$\pm$0.3 & 1.3$^{+1.1}_{-0.6}$ & 0.51$\pm$0.05 & 5.0 & (5.4$\pm$0.54)$\times10^{-3}$ & 4.9$\pm$0.49 & $\sim$2100 & (1), (20) \\ 
GRB 060505 & 0.089 & 0.175 & (0.0) & 0.209 & 0.63$\pm$0.01 & 1.91$\pm$0.2 & (0.8) & $\cdots$ & (3.9$\pm$0.9)$\times10^{-3}$ &120$\pm$12 & $\sim$4 & (1), (21) \\
GRB 060614 & 0.125 & 0.313 & 0.5$\pm$0.4 & 0.068 & 0.11$\pm$0.03 & 2.0$^{+0.3}_{-0.2}$ & 0.8$\pm$0.2 & 4.64 & 0.22$\pm$0.09 & 55$\pm$45 &108.7 & (3), (21) \\ 
GRB 060912A & 0.937 & 0.420	& (0.0) & 1.436 & 0.5$\pm0.3$ & 1.01$\pm$0.06 & 0.6$\pm$0.2 & 3.3 & [0.80 - 1.42] & $>$211 & 5.0 & (4), (17) \\ 
GRB 061021 & 0.3463 & 0.452	& 0.6$\pm$0.2 & 0.185 & $<$ 0.10 & 0.97$\pm$0.05 & 1.02$\pm$0.06 & 3.63 & $\cdots$ & $\cdots$ & 46.2 & (3) \\ 
GRB 061110A & 0.758 & 0.494	& (0.0) & $<$0.10 & $<$ 0.10 & 1.1$\pm$0.2 & 0.4$\pm$0.7 & 3.68 & [0.35 - 0.97] & $>$145 & 40.7 & (3), (17) \\ 
GRB 061210 & 0.4095 & 0.339 & (0.0) & 0.489 & $\cdots$ & 1.67$\pm$0.85 & (0.8) & $\cdots$ & [0.10 - 0.33] & $>$105 & 85.3 & (17) \\
GRB 070419A & 0.97 & 0.24 & $<10$ &0.081 &$<$0.8 & 0.56$\pm$0.0 & (0.8) & $\cdots$ & [0.20 - 0.87] & $<$69 & 115.6 & (5), (17) \\
GRB 071112C & 0.823 & 0.852	& $<$5 & 0.203 & 0.20$^{+0.05}_{-0.04}$ & 1.43$\pm$0.05 & 0.8$^{+0.5}_{-0.4}$ & 3.0 & $\cdots$ & $\cdots$ & 15 & (4), (17) \\ 
GRB 081007 & 0.5295 & 0.143 & 0.97$^{+6.9}_{-0.97}$ & 0.196 & 0.36$^{+0.06}_{-0.04}$ & 1.23$\pm$0.05 & 0.99$^{+0.88}_{-0.43}$ & 4.5 & 0.18$\pm$0.02 & 61$\pm$15 & 10 & (7), (22) \\
GRB 090417B & 0.345 & 0.14 & 22$\pm$3 & 0.083 & 0.8$\pm0.1$ & 1.44$\pm$0.07 & 1.3$\pm$0.2 & 3.54 & [0.17 - 0.35] & $>$70 & $>$260  & (6), (17) \\ 
GRB 090814A & 0.696 & 0.461	& (0.0) & 0.15 & $<$0.2 & 1.0$\pm$0.2 & (0.8) & 3.5 & [0.21 - 0.58] & $<$114 & 80 &(7), (17) \\ 
GRB 100316D & 0.059 & 0.82 & (0.0) & 0.088 & 2.6 & 1.34$\pm$0.07 & 0.5$\pm$0.5 & $\cdots$ & (6.9$\pm$1.7)$\times10^{-3}$ & 20$\pm$10 & 292.8 & (8), (23) \\ 
GRB 100418A & 0.6235 & 0.584 & (0.0) & 0.623 & 0.0 & 1.42$\pm$0.09 & 0.9$\pm$0.3 & 4.82 & [0.06 - 0.15] & $<$50 & 7.0 & (9) (17) \\ 
GRB 101225A & 0.847 & 0.928 & (0.0) & 0.311 & 0.75 & $\cdots$ & (0.8) & 4.65 & [0.68 - 1.2] & $<$98 & 1088 & (12), (17) \\
GRB 110106B & 0.618 & 0.23 & (0.0) & 0.032 & $\cdots$ & 1.35$\pm$0.06 & 1.32$^{+0.67}_{-0.32}$ & 4.04 & 0.73$\pm$0.07 & 194$\pm$56 & 24.8 & (24) \\
GRB 120422A & 0.283 & 0.372 & (0.0) & 1.241 & 0.0 & 1.3$\pm$0.3 & 0.4$\pm$0.4 & 5.07 & [0.016 - 0.032] & $<$72 & 5.35 & (10), (25) \\ 
GRB 120714B & 0.3984 & 0.187 & (0.0) & 0.077 & $\cdots$ & 1.89$\pm$0.02 & (0.8) & $\cdots$ & 0.08$\pm$0.02 & 69$\pm$43 & 159 & (17) \\ 
GRB 120722A & 0.9586 & 0.298 &350$^{+230}_{-170}$ & 0.555 & $\cdots$ & 1.2$\pm$0.4 & 1.2$\pm$1.2 & $\cdots$ & [0.51 - 1.22] & $<$88 & 42.4 & (17) \\ 
GRB 120729A & 0.8 & 1.4 & (0.0) & 0.112 & $\cdots$ & 2.8$\pm$0.2 & 0.8 $\pm$0.3 & 3.9 & [0.80 - 2.0 ] & $>$160 & 71.5 & (17) \\ 
\hline
\end{tabular}
}
Note: for A$_V$ values: (1)~ \citet{sav09}; (2) \citet{pel06}; (3) \citet{zaf11}; 
(4) \citet{sch12}; (5) \citet{kan10}; (6) \citet{per13}; (7) \citet{gre10}; (8) \citet{sta10};
 (9) \citet{mar11}; (10) \citet{can13} ; (11) \citet{mic12}; (12) \citet{cam11};
for E$_{\rm iso}$ \& E$_{\rm p,i}$ values: (13) \citet{pia99}; (14) \citet{ula05}; 
(15) \citet{ama09}; (17) in this work; (18) \citet{cab08}; (19) \citet{sak11}; 
(20) \citet{cam06}; (21) \citet{ama07}; (22) \citet{bis08}; (23) \citet{sta11}; 
(24) \citet{bha11}; (25) \citet{mel12}
\end{table*}

\section{Statistical Properties}
\label{sec_properties}

\subsection{The redshift distribution}

\begin{figure}
\begin{center}
\includegraphics[width=8cm]{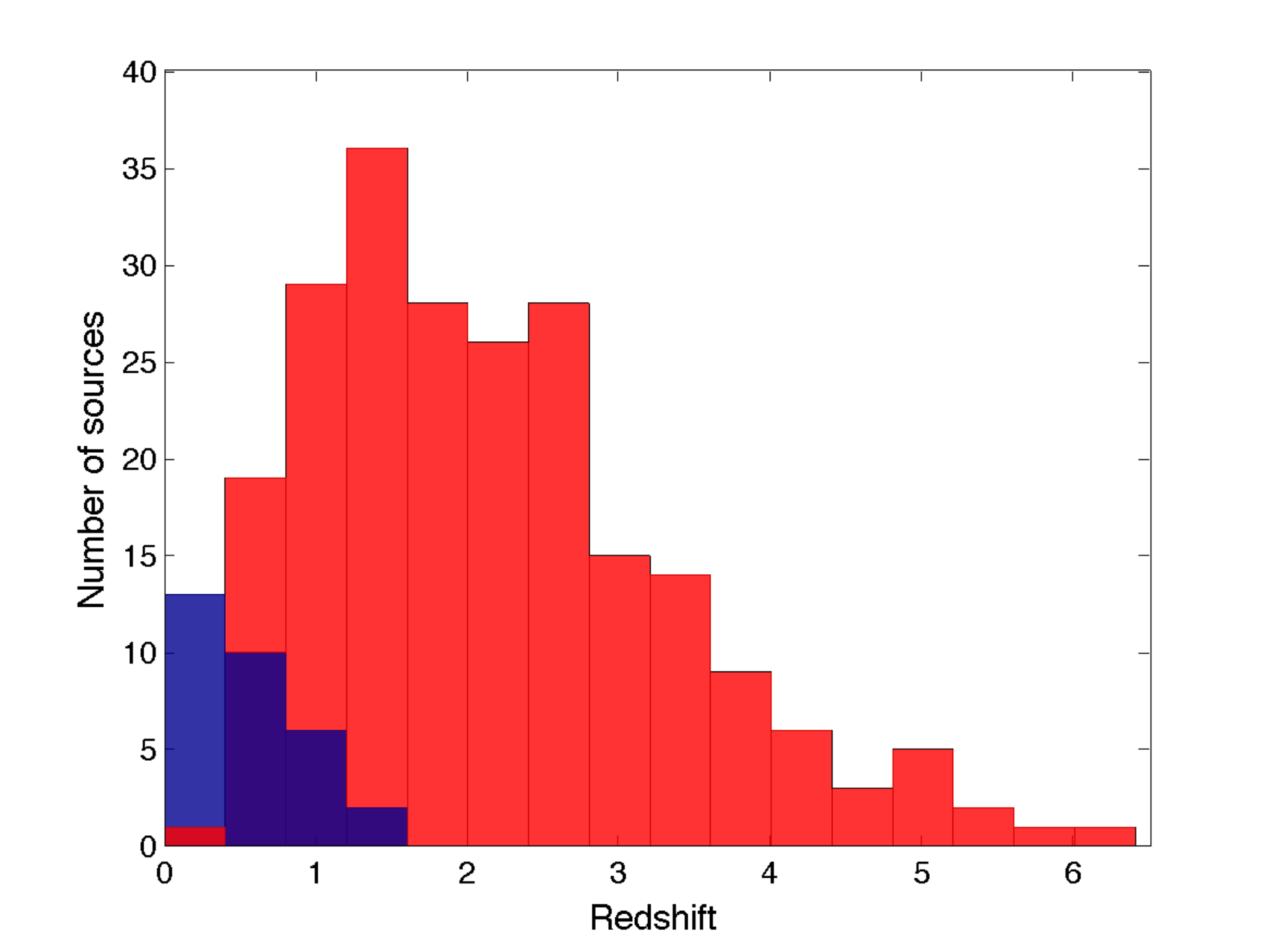}
\includegraphics[width=8cm]{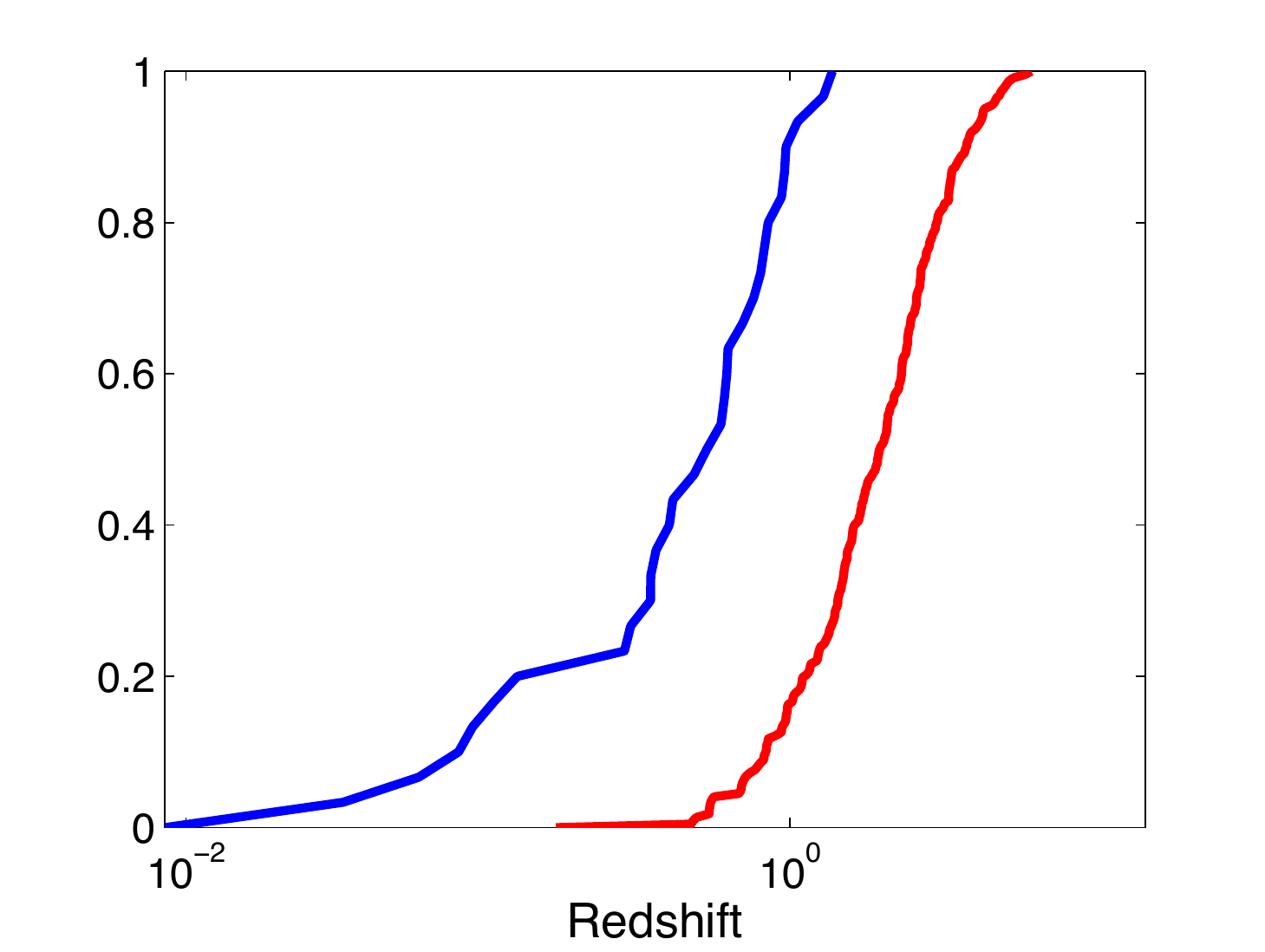}
\caption{Top: Redshift distribution of LLA GRBs (blue) compared to the normal lGRBs (red). Bottom: Cumulative distribution of the same samples.\label{fig_distri_z}}
\end{center}
\end{figure}

The redshift distributions of the LLA GRB sample is shown in Fig. \ref{fig_distri_z}, together with the distribution of all lGRBs.

A simple examination of  Fig. \ref{fig_distri_z} shows that LLA GRBs are closer than normal lGRBs whose distribution peaks on average at z = 2.2 \citep[\textit{e.g.}][]{jak06,cow13}. 
Table \ref{table_fit_z}  displays the main parameters of the two distributions. We performed a Kolmogorov-Smirnov test on the two data sets, which shows that the probability for the two distributions to be based on the same population is $1.1 \times 10^{-14}$, hence rejecting this hypothesis.

\begin{table}
\centering
  \caption{Statistical parameters of the cumulative redshift distributions.\label{table_fit_z}}
   \begin{tabular}{lrr}
  \hline
  Parameter& LLA GRBs & All lGRBs\\
    \hline
mean&0.55&2.20\\
median&0.53&1.98\\
standard deviation&0.38&1.19\\
\hline
\end{tabular}
\end{table}

We check now whether the difference between the two redshift distribution is intrinsic or due to a selection bias.

\begin{figure}
\begin{center}
\includegraphics[width=8cm]{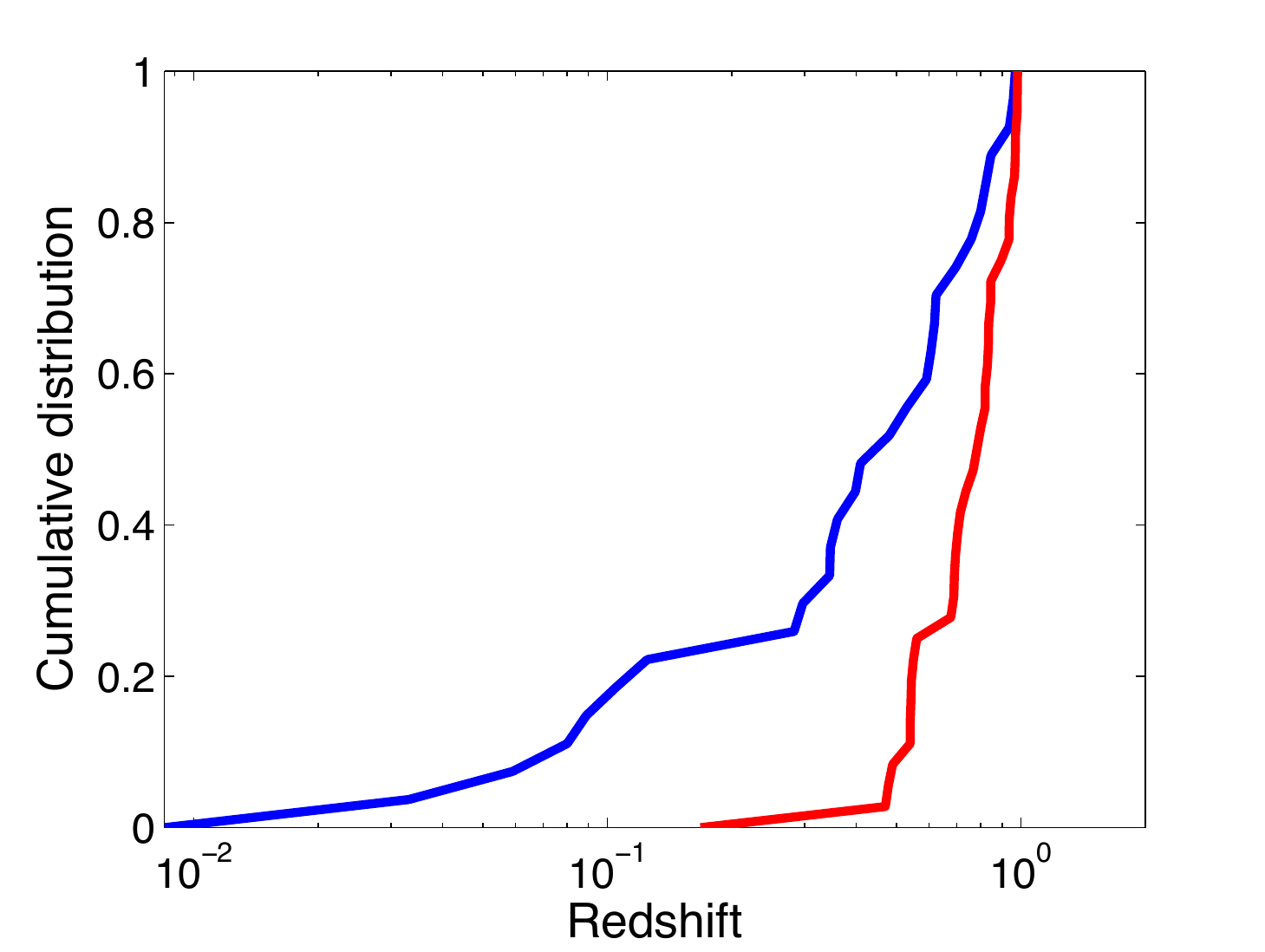}
\caption{Cumulative distribution of the redshift of LLA GRBs (blue) compared to all GRBs (red) for redshifts z$<$1.\label{fig_distri_z<1}}
\end{center}
\end{figure}

Faint events are more difficult to detect than brighter ones. Furthermore, the measure of the redshift implies that the optical afterglow is bright enough for spectroscopic observations to be performed. As a matter of consequence, LLA GRBs are plagued by a detection bias that prevent them to be detected at large distance. From their flux, we estimate that the faintest of the LLA GRBs present in our sample can be detected up to a distance of z = 1. Because regular lGRBs can be detected up to z = 8.2 \citep{tan09}, we also consider that the sample of lGRBs is complete for $z < 1$, thus removing the detection bias. We have recomputed the cumulative redshift distributions for this sub-sample (see Fig. \ref{fig_distri_z<1}). The difference is still large, and from a Kolmogorov-Smirnov test the probability that the two distributions are drawn from the same population is $9.4 \times 10^{-4}$. We thus conclude that the LLA GRB population is different from the "classical" lGRBs one.

\subsection{Absorption and Extinction}

For consistency, we first checked that our distribution for the Milky Way values of $\text{A}_{V,Gal}$ and $\text{N}_{H,Gal}$ (i.e. the optical extinction and X-ray absorption parameters) is consistent with the whole sample of normal long GRBs. While the X-ray absorption has little effect on our sample since we use the flux in the 2.0-10.0 keV band, where absorption can be neglected \citep{mor83}, it is well known that the optical extinction can bias a distribution \citep[for instance the well known problem of dark bursts, e.g.][]{jak04}.

\begin{figure}
\begin{center}
\includegraphics[width=8cm]{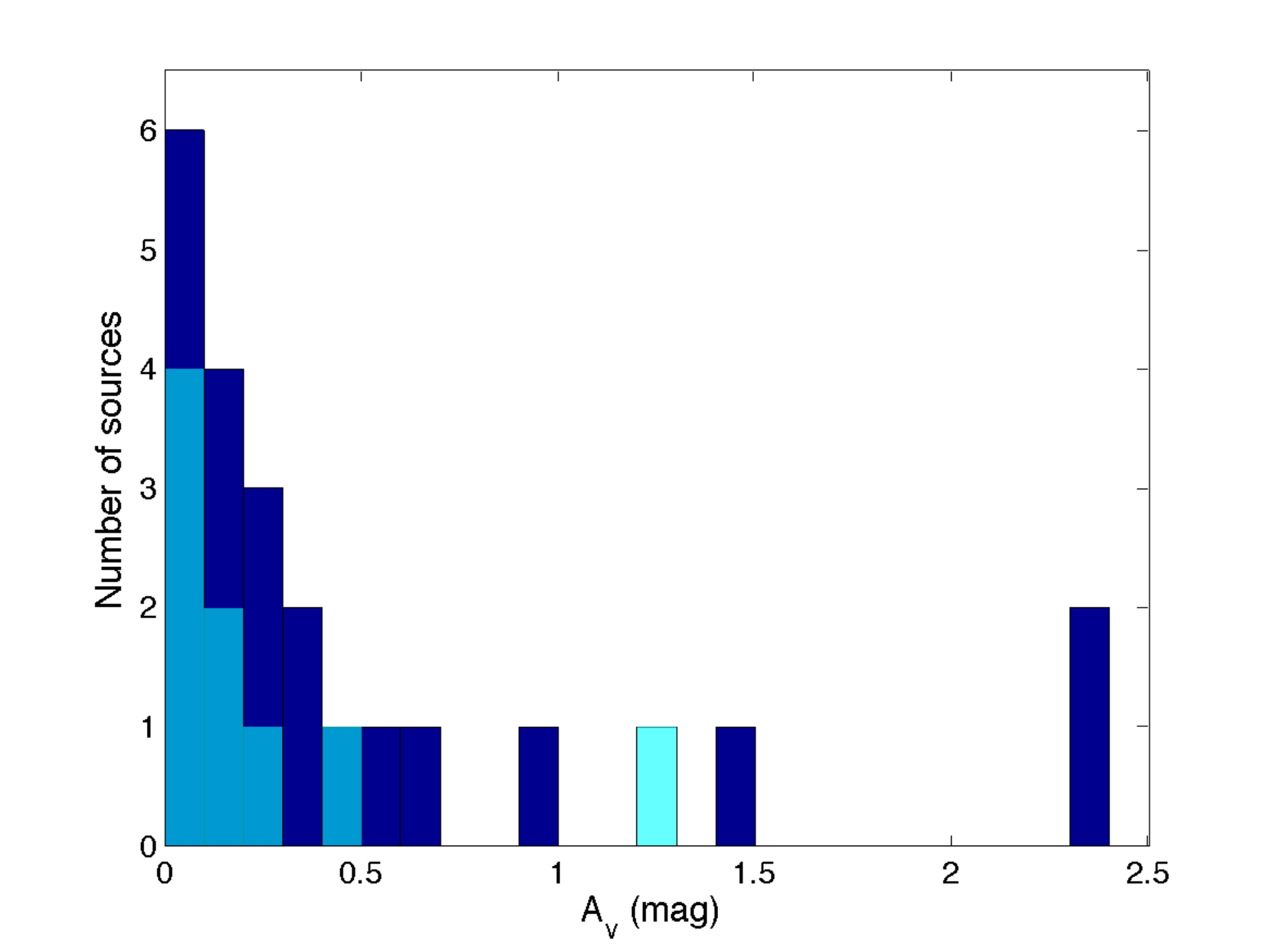}
\caption{Histogram of the optical extinction $\text{A}_V$ in the Galaxy for LLA GRBs. GRB-SN associations are represented by cyan bars while other LLA GRBs are shown with blue bars.\label{fig_AV}}
\end{center}
\end{figure}

The optical extinction was calculated using the NASA/IPAC extragalactic database\footnote{http://ned.ipac.caltech.edu/forms/calculator.html} for the  Landolt V band measured by \citet{sch98} for all bursts but GRB 060904B and GRB 061110A. These two bursts are seen in projection on the galactic disk where the measures of \citet{sch98} are highly variable with the position. For these two events, we rely on the most accurate measurements of \citet{sch12} and \citet{zaf11} respectively. The results are reported in Fig. \ref{fig_AV}. One can note that the A$_V$ values obtained for LLA GRBs statistically are not very larger compare to the one for normal lGRBs (A$_V$ < 2 for 87\% of normal lGRBs, \citep{cov13}). In the following, we consider that the gas and dust in the Galaxy have not introduced a bias in LLA GRB sample.

\begin{figure}
\begin{center}
\includegraphics[width=8cm]{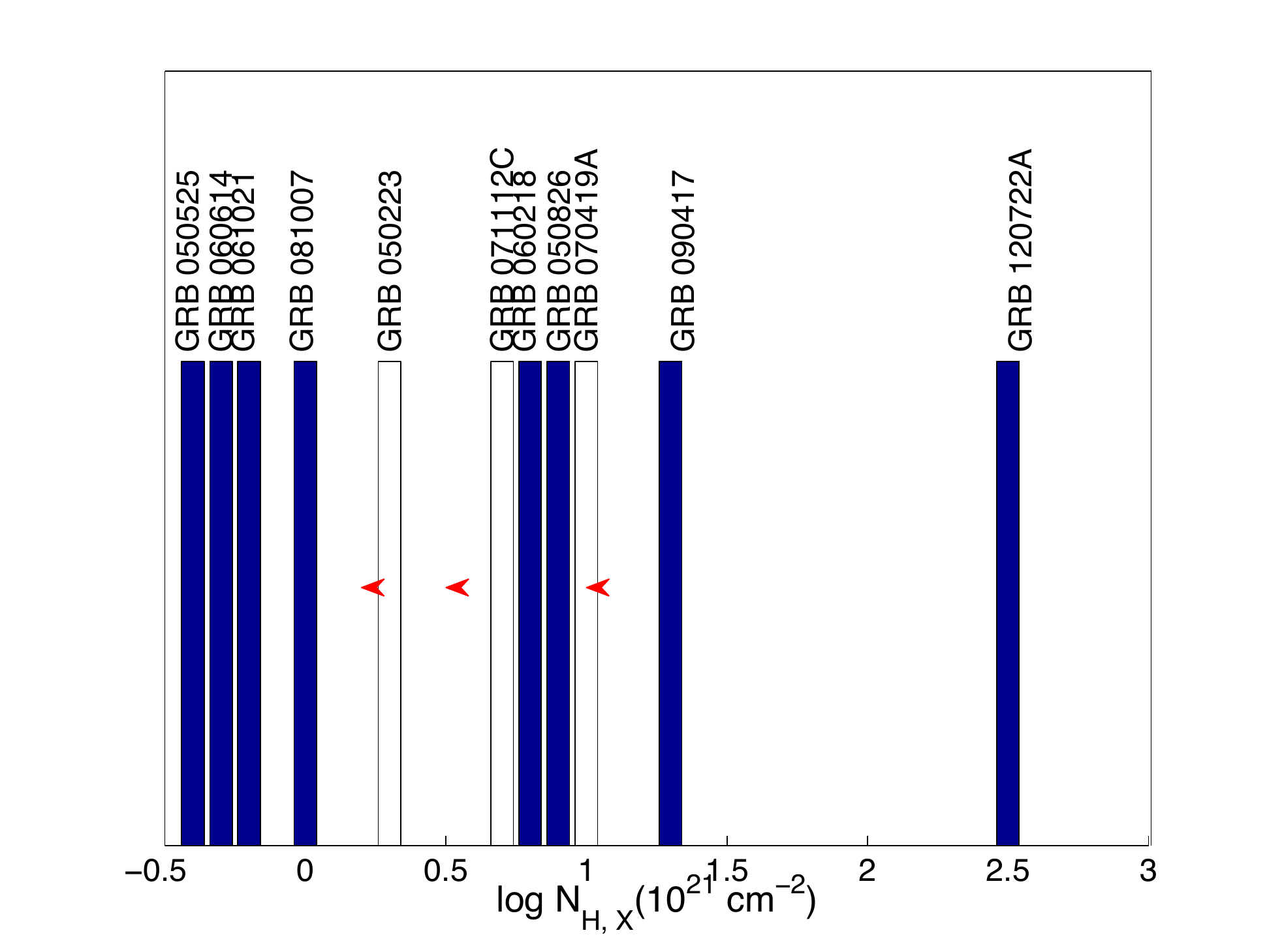}
\caption{The HI column density in the host galaxy, $\text{N}_{H,X}$ for 11 LLA GRBs. The blue bars are fitted values while the white ones with red arrows are upper limits. \label{fig_distri_NH}}
\end{center}
\end{figure}

The intrinsic hydrogen column density $\text{N}_H$ can be linked to the host properties \citep{rei02}, thus we also investigated on this. The intrinsic host absorptions for the LLA GRBs are mostly compatible with little or no intrinsic absorption. We see that for the sources with a non zero $\text{N}_{H,X}$ (Fig. \ref{fig_distri_NH}), the absorption of the host galaxy is on average a factor 10 larger than in the Milky Way, as already noted by \citet{sta13}. At low redshift this effect was attributed to the gas in the host galaxy.

\subsection{Decay and spectral index}

\begin{figure*}
\begin{center}
\includegraphics[width=8cm]{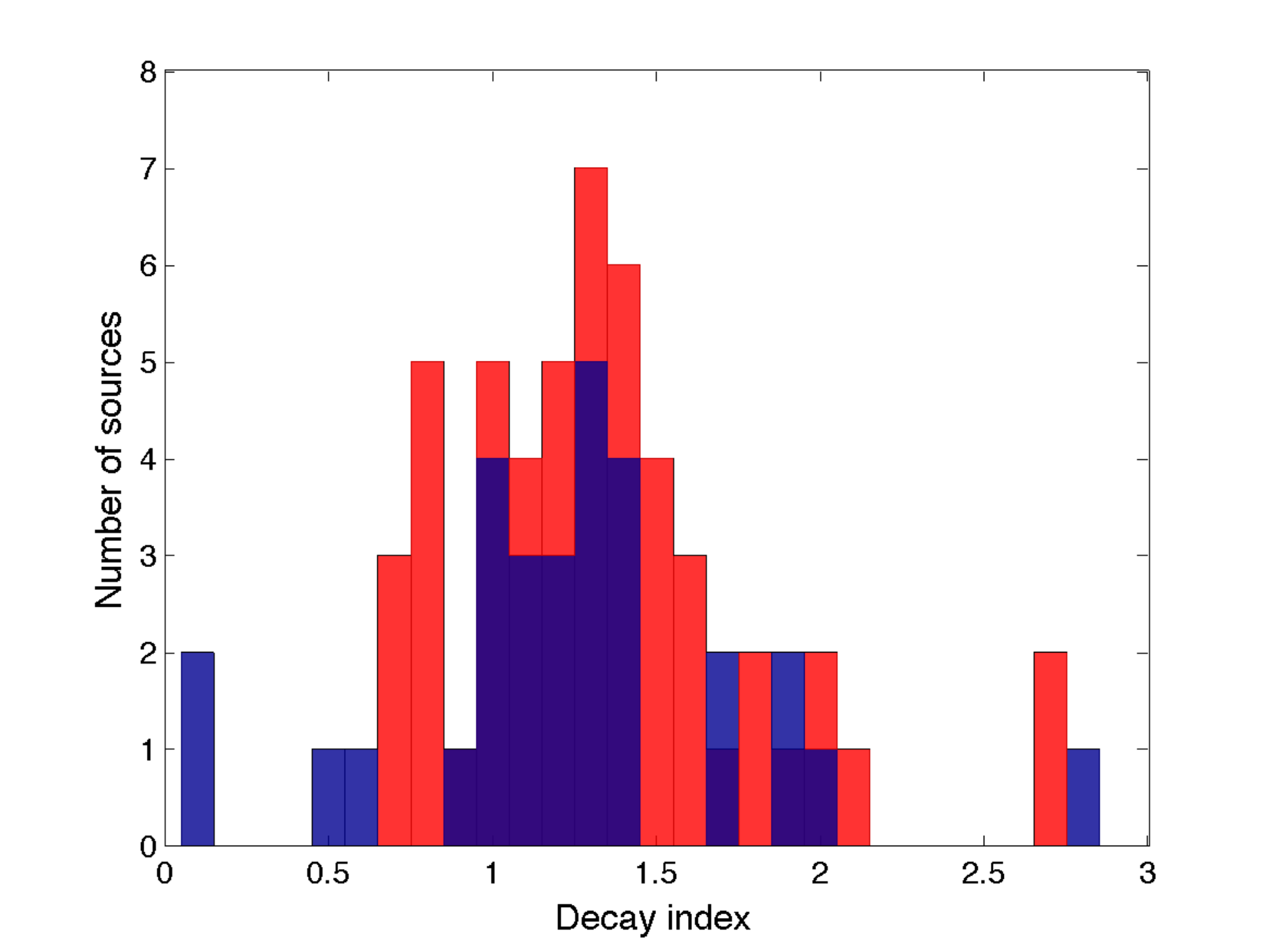}
\includegraphics[width=8cm]{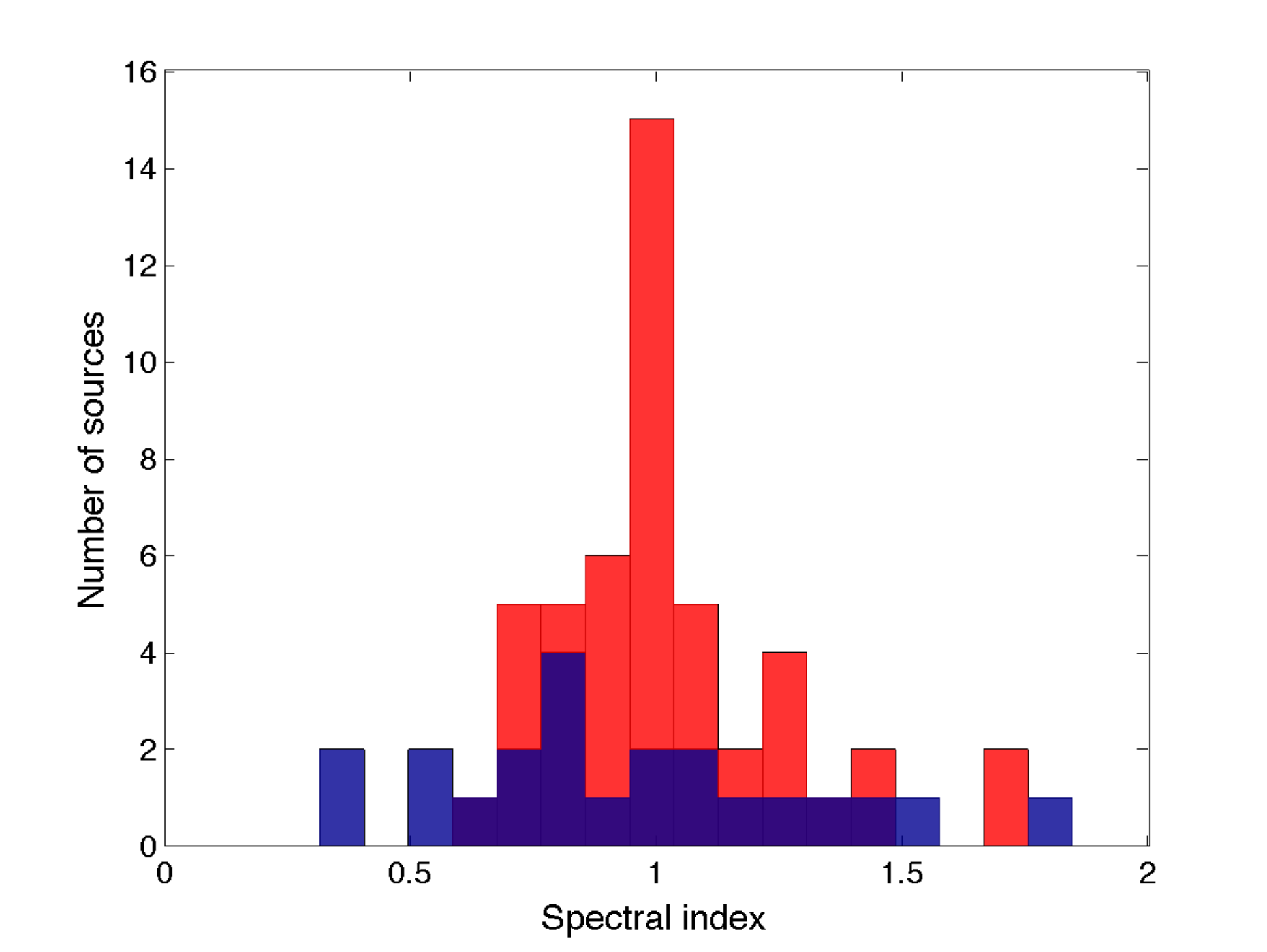}
\caption{Distribution of the decay index (left) and spectral index (right) for the LLA GRBs sample (blue) compared to the reference sample (red).\label{decay_index}}
\end{center}
\end{figure*}

\begin{figure*}
\begin{center}
\includegraphics[width=16cm]{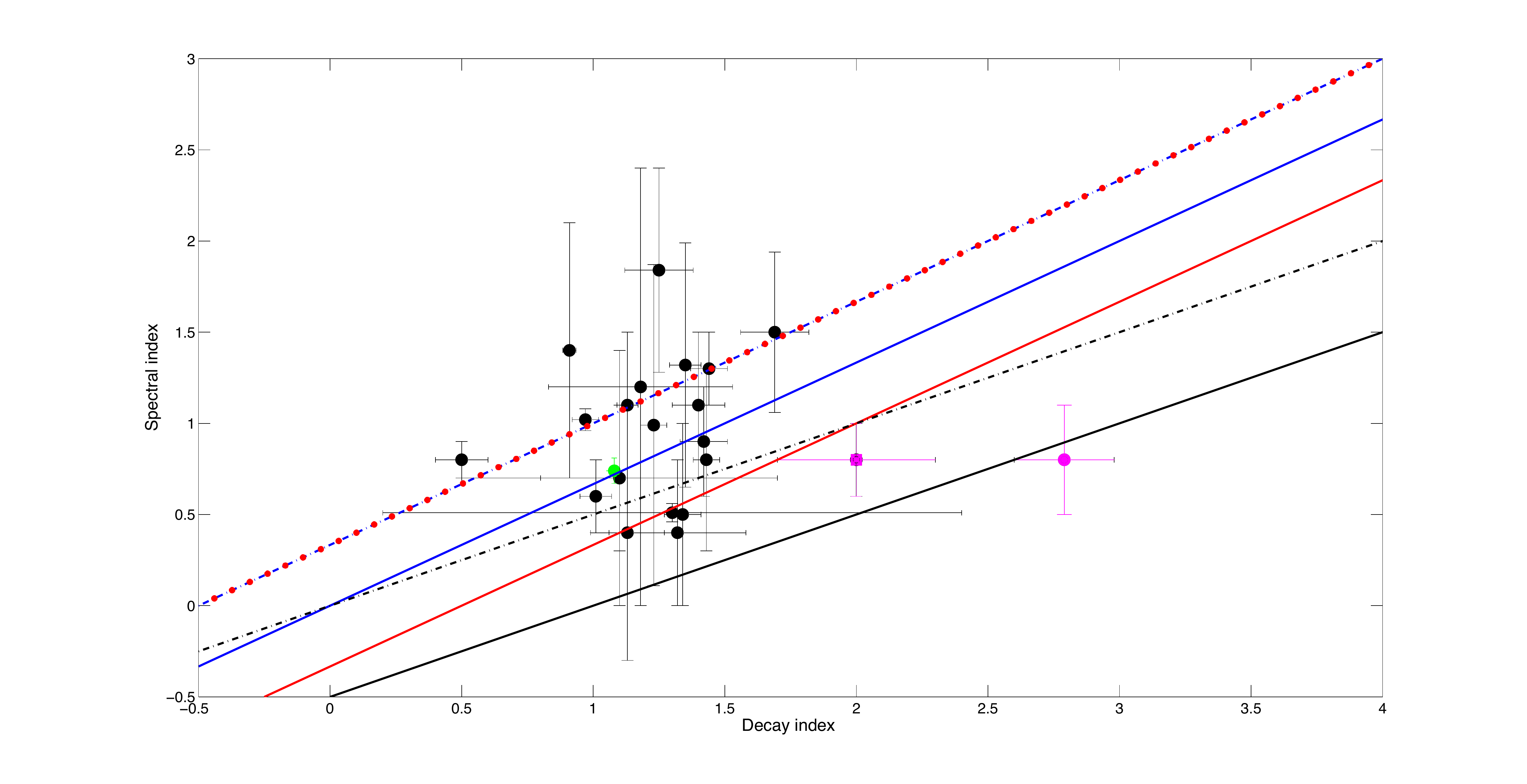}
\caption{X-ray decay index versus spectral index of LLA GRBs. The purple filled circle and square represent GRB120729A and GRB 060614 respectively. The green dot represent GRB 120729A before the break at $t_b = 8.1$ ks. All closure relations, indicated by the lines, are computed for $p > 2$ in the slow cooling phase. Solid and dash-dotted lines stand for $\nu_m < \nu < \nu_c$ and $ \nu_c < \nu$ respectively. Blue, red and black lines stand for ISM, wind medium, and jet effect respectively. \label{closure}}
\end{center}
\end{figure*}

The distributions of the temporal decays and spectral index for the LLA GRB sample are displayed in Fig. \ref{decay_index}. We used a reference sample of bursts listed in \citet{gen08} and not members of the LLA GRB subclass. Note that the decay index of GRB 060607A reported in this last article is incorrect and is not considered in the comparison. The two samples are very similar, as indicated by a K-S test (p = 0.80 and p = 0.08 for the decay and the spectral indices respectively). We thus conclude that the two samples have similar spectral and temporal properties. This can also be seen when considering the closure relations \citep{mes98, sar98, sar99a, che00, zha08} to investigate the burst geometry, the fireball microphysics, its cooling state and the surrounding medium (presented in Fig. \ref{closure}). These are very similar to the ones obtain from BeppoSAX, XMM-Newton or Chandra \citep{dep06, gen06} for long bursts. We note however two peculiar events:

\begin{enumerate}
\item GRB 120729A: The pre-jet break closure relations are rejected for this event. We can thus identify the break presents at $t_b = 8.1$ ks in the light curve as the jet break, obtaining the positions of the specific frequencies and the value of p ($\nu_m < \nu_{XRT} < \nu_c$, $p = 2.8 \pm 0.2$). The pre-jet break decay properties are in agreement with this (green point in Fig. \ref{closure}).

\item GRB 060614: This event would be compatible with another jet effect, with $p = 2.25 \pm 0.05$. However, the errors bars are large enough to accommodate some non jetted closure relations. We thus cannot conclude firmly on the jet hypothesis for this source based on the closure relations alone.
\end{enumerate}

\subsection{Prompt phase}

We also investigated the prompt properties of the LLA GRBs. For this purpose, as the BAT bandwidth is narrow, we used whenever possible the data from  {\em Fermi} GBM. For events seen by Konus-Wind or BeppoSAX, we used previously published results. About half of the events have a firm measurement of the prompt parameters, the other half presenting upper and lower limits. We corrected for the cosmological redshift the values of E$_p$, obtaining the intrinsic E$_{p,i}$ values. These values cluster broadly within the 40-200 keV range. 

We note however that there is a lack of bright events also in the prompt phase. Taking into account the median redshift of LLA GRBs, the E$_{p,i}$ values, the intrinsic scatter of the Amati relation, and the properties of the BAT instrument, one should expect to detect bursts up to E$_{\rm iso} = 3 \times 10^{53}$ ergs, at least one order of magnitude larger than the brightest measurements listed in Table \ref{table_sample}. We thus conclude that this effect is an evidence that LLA GRBs are intrinsically less energetic, both during the prompt and the afterglow phases, compared to normal lGRBs.


\section{Discussion}
\label{sec_discu}

\subsection{Distance of the sample}

As noted by previous authors \citep[e.g.][]{watson04, gue07, dai07, imerito}, faint GRBs cannot be detected at large distance, and, by definition, all LLA GRBs have a faint luminosity afterglow. We are thus missing distant LLA GRBs, as one could expect from the redshift distribution.

On the other hand, can we consider that all bursts with a faint afterglow are LLA GRBs, and thus that our sample is not contaminated by some normal lGRBs? We assume this is not the case based on our selection criteria, which allows to discriminate regular nearby lGRBs such as GRB 030329 (which is not part of our sample, and in fact a normal lGRB).

Is this distribution of redshift  biased? The volume of the Universe at low redshift is very small, thus allowing for few events to occurs: this could explain the lack of normal lGRBs at redshift lower than 0.3. We note however that this argument also apply to LLA GRBs, and thus that if the two populations were similar in their redshift distribution, we should see the same proportion of bursts located between $0 < z < 0.3$ and $0.5 < z < 1.0$ for LLA GRBs and normal lGRBs. As can be seen in Fig. \ref{fig_distri_z} this is not the case. We thus conclude that, if our sample is contaminated, the proportion of normal lGRBs is not large enough to prevent the main properties of this group to be apparent, and that LLA GRBs are events closer than normal lGRBs.

\subsection{Geometry and environment of the burst}

Most of the sources can be explained by both a wind environment and a constant ISM. As shown below, many of these sources are associated with SNe (see Table \ref{table_SNe}). This association would point towards a wind environment \citep{che04}. However, as shown by \citet{gen07}, the termination shock can lie very nearby to the star, and we cannot conclude firmly on the surrounding medium. 

One source deserves a more careful study: GRB 120729A. This event can be accounted for by the closure relation of a jet. There is also a hint of achromaticity, as a break is seen both in X-ray and in optical around the same time \citep{mas12, dav12b}. This supports the interpretation of a jet effect. 

The opening angle is given by \citet{lev05} who extended the work of \citet{sari99b} to account for the radiation efficiency of the prompt phase:
\begin{equation}
\label{eq_opening}
\theta (t_{b}, E_{iso} ) = 0.161\left(\frac{t_{b,d}}{1+z}\right)^{3/8}n^{1/8}\left(\frac{\eta_\gamma}{E_{iso,52}} \right)^{1/8}\text{,}
\end{equation}  
where the standard values for the number density of the medium $n=1\text{cm}^{-3}$ and the radiative efficiency $\eta_\gamma=0.2$ are used. We get $\theta = 2.7^\circ$. 

From the post jet-break part of the light curve, we derive $p = 2.8 \pm 0.2$. This value is not compatible with {\it both} the spectral and temporal decay indexes (0.74$\pm$0.072 and 1.08$\pm$0.03 respectively) of the pre-break part of the light curve. Only the spectral index is marginally consistent with this value of p, assuming $ \nu_m < \nu_x < \nu_c$ and an ISM. The temporal decay is too flat (we expect a value of at least 1.5). In order to reconcile all of these facts, we need to involve some late time energy injection to flatten the light-curve \citep{has14}. This energy injection needs to be present during the pre-break part of the light curve, but should stop during the post-break part. We  note that the sampling of the X-ray light curve is not good during the jet break and allows for some non-simultaneity.  

A similar argument can be drawn for GRB 060614, which may also be compatible with a jet, according to the closure relations. This burst also displays an achromatic break (around 36.6 ks) in X-ray, optical and UV \citep{man07}. However, before the jet, this burst features a plateau phase and not a standard afterglow. This is somewhat unusual for a GRB, and would request energy injection fine tuned in order to stop at the moment of the jet break.  We note in addition that GRB 060614 has been proposed to be a short GRB with an extended emission \citep{060614_short}, making this event clearly odd in our sample. The explanation of why the energy injection would stop at the same time than the jet break is beyond the scope of this paper. In any cases, if we assume the presence of a jet, the corresponding jet opening angle is $6.3^\circ$.

Several authors \citep[e.g.]{yam03, ram05, dai07} have tried to explain the properties of some of these events based on the jet properties (aperture angle, viewing angle). From our findings, when we do have a measurement of the jet aperture it is similar to the one of normal lGRBs \citet[$\theta = 4.7 \pm 2.3^\circ$,]{ghi12}, and there is no hint of large off-axis viewing. The typical LLA GRB should have a jet not very different from the one of normal lGRBs.

\subsection{Microphysics of the fireball}

The spectral and temporal properties, once merged into the closure relations (see Fig. \ref{closure}), can indicate the position of the cooling frequency and thus give some insight into the parameters of the fireball. For LLA GRBs, we have two possibilities: either we cannot conclude, or the X-ray band is located {\it below} the cooling frequency. In the former case, this is just due to the uncertainties of the measurement. In the latter, this is not common: indeed, most late GRB afterglows are compatible with the X-ray band located above the cooling frequency \citep{gen06, dep06}. We insist here on the fact that this measure is time dependent, as shown in \citet{gen06}, and that the comparison need to be done with consistent data.

In the case of an homogeneous interstellar medium (ISM), the formula of the cooling frequency is \citep{pan00} :
\begin{equation}
\label{eq_nuc1}
\nu_c = 3.7 \times 10^{14} E_{53}^{-1/2} n^{-1} (Y+1)^{-2} \epsilon_{B,-2}^{-3/2} T_d^{-1/2} {\rm Hz}\text{,}
\end{equation}
where $E_{53}$ is the isotropic energy in units of $10^{53}$ ergs, $\epsilon_{B,-2}$ is units of $10^{-2}$, n is the number density of the medium in the units of $cm^{-3}$ , Y is the Compton parameter, $\epsilon_{B,-2}$ is the fraction of internal energy of magnetic field, $T_d$ is the time expressed in days after the burst.

Instead in the case of a wind medium, the cooling frequency reads
\begin{equation}
\label{eq_nuc2}
\nu_c = 3.5 \times 10^{14} E_{53}^{1/2} A_*^{-2} (Y+1)^{-2} \epsilon_{B,-2}^{-3/2} T_d^{1/2} {\rm Hz}\text{,}
\end{equation}
where $A_*$ is the number density in the wind.

For those bursts where we can conclude on the position of the cooling frequency, we can then insert the numbers to have an idea of the constraints on the model. We start by assuming that the fireball expands in the ISM. The XRT band ranges from $7.2 \times 10^{16}$ Hz to $2.4 \times 10^{18}$ Hz respectively. We however assume that $\nu_c$ is above $3.7 \times 10^{18}$ Hz (i.e. slightly above the X-ray band) for simplicity. Equation \ref{eq_nuc1} simplifies to :
\begin{equation}
\label{eq_nuc3}
10^{-4} E_{53}^{-1/2} \epsilon_{B,-2}^{-3/2} < 1\text{,}
\end{equation}
when assuming the standard density $n=1 \text{cm}^{-3}$, the Compton parameter $Y \ll 1$ and considering the observation time of 1 day. Taking the lowest $E_{53}$ measured (in order to have the stringent constraint), i.e. $10^{-5}$ (value of $E_{53}$ for GRB 980425), we obtain: $\epsilon_{B,-2} > 0.1$, which is not really constraining, as typical values of $\epsilon_{B,-2}$ should be of the order of 1 for lGRBs. 

The situation is similar when assuming a wind medium, for which Eq. \ref{eq_nuc2} implies : 
\begin{equation}
\label{eq_nuc4}
10^{-4} E_{53}^{1/2} \epsilon_{B,-2}^{-3/2} < 1\text{,}
\end{equation}
when assuming a standard density and a Compton parameter, and considering the observation time. Using the same method, but this time using the largest value of $E_{53}$ (again in order to have the stringent constraint), we obtain $\epsilon_{B,-2} > 0.0012$. Again, the magnetic energy of the fireball seems not to explain the unusual position of the cooling frequency.

We thus conclude that, under both hypotheses, the uncommon position of the cooling frequency for LLA GRBs compared to normal lGRBs is due to the small energy of the fireball rather than the magnetic energy of the fireball.

\subsection{Prompt properties of LLA GRBs}

\begin{figure*}
\begin{center}
\includegraphics[width=8cm]{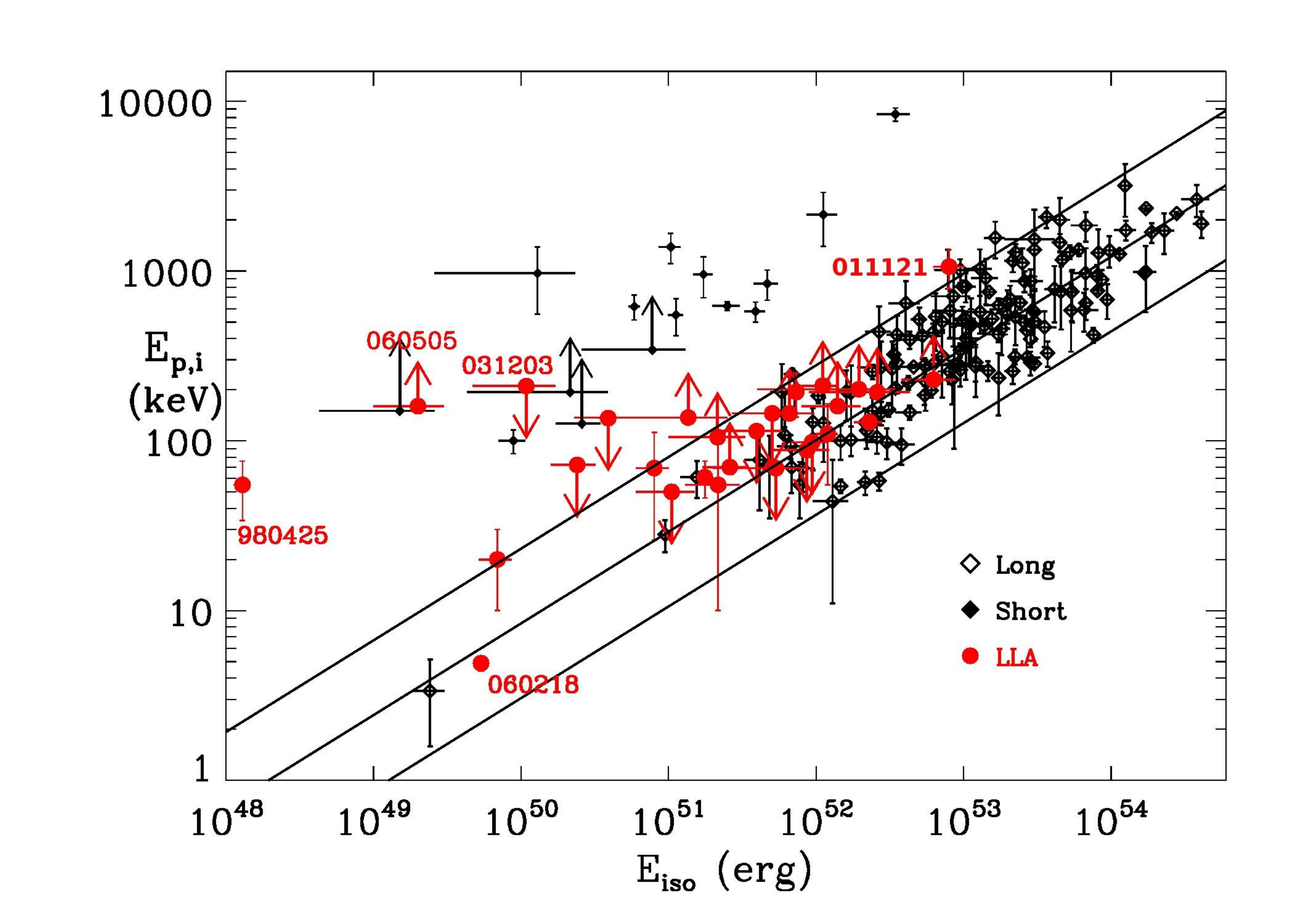}
\includegraphics[width=8cm]{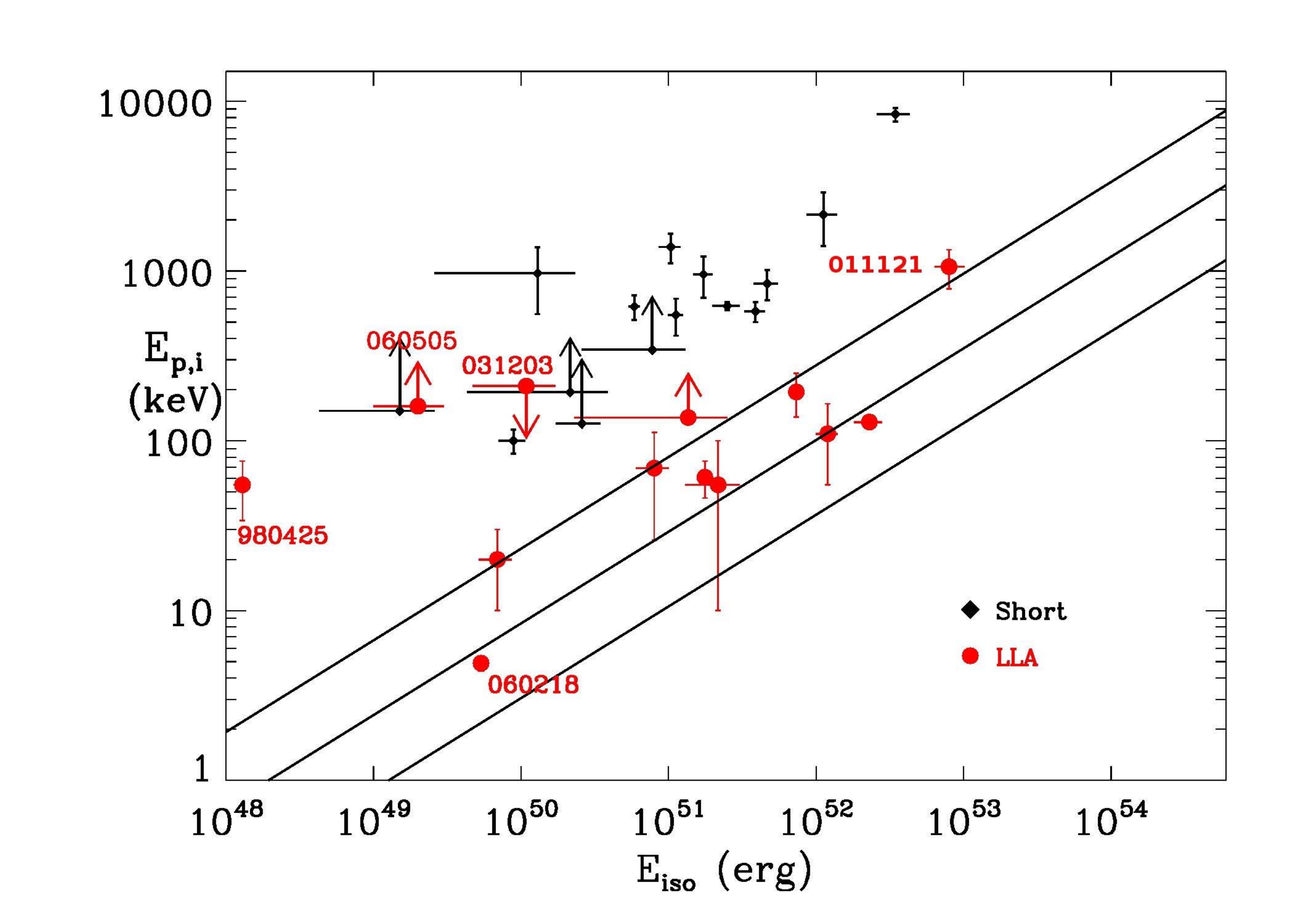}
\caption{Location in the E$_{p,i}$-E$_{\rm iso}$ plane of LLA GRBs sample. Left, comparison of  both short and long GRBs. Right, compared to short events, with firm measurements and non-compatible lower limits on E$_{p,i}$ only. \label{fig_amati}}
\end{center}
\end{figure*}

In Fig. \ref{fig_amati}, we clearly see that all the outliers of the Amati relation belong to the LLA GRB sample. Several explanations have been proposed to explain these events \citep[see][and reference therein for details]{ama06, ama07}: GRB 060505 may be a short GRB (as its location in the E$_{p,i}$ - E$_{\rm iso}$ plane may suggest); the E$_{p,i}$ value of GRB 061021 refers to the first hard pulse, while a soft tail is present in this burst (so the true E$_{p,i}$ may be lower); GRB 031203 may be much softer than measured by INTEGRAL/ISGRI as supported by dust echo measured by XMM. 

We can indeed see that the outliers are all located in the left part of the E$_{p,i}$ - E$_{\rm iso}$ plane. In this part of the diagram, the usual gamma-ray instruments are not well suited to measure the prompt properties. For instance, the BAT measurements of GRB 060218 \citep{sak06} would have make this event more similar to GRB 980425, i.e. a clear outliers. It is its simultaneous observation by BAT and XRT that makes it compatible with the Amati relation. One may thus imagine that this conclusion may hold for all outliers. We note however that the prompt phase of a GRB usually shows a hard-to-soft spectral variation \citep[e.g.][]{mes06} and that the prompt emission of GRB 060218 lasted significantly longer than other bursts: it is not sure that a measurement consistent with the ones done at the BeppoSAX epoch (i.e. time averaged over the complete prompt emission) would lead to a similar conclusion.

On the other hand, GRB 980425, GRB 060505 and (marginally) GRB 050826 are not compatible at all with the Amati relation. If we assume that these measurements are correct, then the best fit relation in the E$_{p,i}$ - E$_{\rm iso}$ plane changes dramatically, being far more flatter. 

A flatter Amati relation has been foreseen as early as 2003 \citep{yam03}, using GRBs seen off-axis. For completeness, we also note that a similar explanation hold in case of the canonball model \citep{dad05}. Being seen off-axis, these events are expected to be less luminous than normal lGRBs {\it even during the afterglow} \citep[see e.g.][]{dal06}. A balance need however to be made between this argument and the fact that the events are detected: once seen off-axis, the luminosity of the burst decrease very fast with the off-axis angle. The simple fact that these events are detected means that they are not at a very large off-axis angle, but should rather be considered slightly off-axis.

Also, as already indicated, there was no a-priori reason why the LLA GRBs have a low E$_{\rm iso}$ (and thus a low prompt luminosity). If it is known that the prompt and afterglow luminosities are linked together \citep{dep06, ghe08, eva09}, this result holds for normal lGRBs and should not considered as obvious for LLA GRBs. We can confirm it also holds for this class of events.

\subsection{LLA GRBs and supernovae}

Nine LLA GRBs are firmly associated to SNe by spectral and photometric optical observations. They are listed in Table \ref{table_SNe}, together with the other solid associations\footnote{While we consider in the following only positive associations, we note that two other sources (GRB 070419A, GRB 100418A) might be associated to SNe. GRB 070419A displays a faint bump in its light curve similar to the one of GRB 980425 \citep{hill07}; A bright host galaxy may prevent to observe the signature of a faint SN associated with GRB 100418A, which can be fainter  than r magnitude -17.2, comparable to the magnitude of the most faintest Ic SNe \citep{nii12}.}.

In our sample, GRB 060505 \citep{hai06} and GRB 060614 \citep{fyn06,del07,gal06} are firmly not associated to SNe. Because of their close distance, any SNe would have been detected, and thus these non associations are significant. This may strengthen the conclusion that these two events are two short bursts, or at least two events not related to a normal colapsar. From our criteria, these events are not short, and thus clearly belong to the LLA GRB subclass. We then can conclude that at least some of these events can be explained by a different kind of progenitor compared to normal lGRBs. 

We also note that a large fraction of the GRB-SNe associations (64\%) belong to the LLA GRBs sample. The positive associations include several well known events, such as GRB 980425/SN1998bw, GRB 031203/SN2003lw, and GRB 060218/SN2006aj. We stress that this can lead to some problems, as GRB 980425/SN1998bw is commonly used as template for light curves and spectra when looking for a SNe within the dataset of a given GRB. As discussed above LLA GRBs progenitors may differ from normal lGRB ones, which also applies to the associated supernovae. It is thus more accurate and safe to use as template GRB 030329/SN2003dh for normal lGRBs.

\begin{table*}
 \centering
  \caption{The table of GRB-SN  associations. 
  The L$_{\rm iso}$ values are calculated from the E$_{\rm iso}$ values which are given in the Table \ref{table_sample} \label{table_SNe}.}
   {\scriptsize
  \begin{tabular}{lllllllll}
  \hline
  GRB & & SN & SN & SN &Host & LLA &L$_{\rm iso}$ of GRB & \\
   name & redshift & identification & name & type& type & GRB&  (10$^{49}$erg s$^{-1}$) & Ref. \\ 
 \hline
GRB 980425 & 0.0085 &spectral&SN1998bw & BL-lc &dwarf spiral & yes & 0.033 & (1) \\
&&& & & (SbcD) & &  \\
GRB 011121 & 0.36 &spectral &SN2001ke & IIn & N/A & yes &387  & (2) \\
GRB 021211& 1.01 &spectral &SN2002lt & $\sim$Ic & N/A & no & 969 & (3), (4), (5) \\
GRB 030329& 0.168 &spectral &SN2003dh & BL-Ic & N/A  & no & 75 & (6), (7), (8) \\
GRB 031203 & 0.105 &spectral&SN2003lw & BL-Ic &Irr & yes & 0.56 & (9), (10) \\
&&& & &Wolf-Rayet & &  \\
GRB 050525 & 0.606 &spectral &SN2005nc & $\sim$Ic& N/A & yes  & 417 & (11), (12) \\
GRB 060218 & 0.0331 &spectral&SN2006aj & BL-Ib/c & dwarf Irr & yes & 0.02  & (13), (14), (15)\\
GRB 081007 & 0.5295 &bump &SN2008hw  & Ic & N/A & yes & 30 & (16), (17) \\
GRB 091127& 0.49 &bump&SN2009nz & BL-Ic & N/A & no & 345 & (18), (19) \\
GRB 100316D & 0.059 &spectral &SN2010bh & BL-Ic & Spiral blue & yes & 0.056 & (20), (21) \\
GRB 101219B & 0.55 &spectral &SN2010ma & Ic & N/A & no & 29 & (22), (23) \\
GRB 120422A & 0.283 &spectral &SN2012bz & Ib/c & N/A & yes & 0.44 & (24), (25) \\
GRB 120714B & 0.3984 &spectral& SN2012eb & I & no identification & yes &0.7 & (26), (27) \\ 
GRB 130215 & 0.597 &spectral &SN2013ez & Ic & N/A & no & 75 & (28), (29) \\
\hline
\end{tabular}
}

Note: for GRB-SN associations: (1)~ \citet{gal98}; (2) \citet{blo02}; (3) \citet{cre02};
(4) \citet{del03}; (5) \citet{vre03}; (6) \citet{gol03}; (7) \citet{kaw03}; (8) \citet{sta03};
(9) \citet{sod03}; (10) \citet{tag04}; (11) \citet{del06}; (12) \citet{blu06}; (13) \citet{cob06}; 
(14) \citet{cam06}; (15) \citet{sod06}; (16) \citet{sod08}; (17) \citet{mar08}; (18) \citet{cob09}; 
(19) \citet{wil09}; (20) \citet{cho10}; (21) \citet{sak10}; (22) \citet{van10}; (23) \citet{spa11};
(24) \citet{mel12}; (25) \citet{bar12}; (26) \citet{klo12}; (27) \citet{cum12}; (28) \citet{deu13};
(29) \citet{you13}; 
\end{table*}

\section{Conclusions}
\label{sec_conc}

As already noted all papers published so far are based on the observed properties of singular faint events, or in some rare cases, two GRBs. GRB 980425-like bursts are faint in their prompt phase and cannot be detected at large distance; in addition they are not compatible with the Amati relation linking E$_{iso}$ to E$_p$. However, this does not extend to all burst with prompt low luminosity that are outliers of the Amati relation. As an example,  GRB 060218 does follow this relation, despite being faint in its prompt phase. In order to have the global answer, one needs to perform the study on a statistically significant sample of events, and this has not been done so far to our knowledge.

Here we have  based our study and selected the sample on the afterglow properties, after the plateau phase, alone. The fact that this made incidentally entering some GRBs with a low-luminosity prompt was therefore not pre-supposed. This approach allows us to build a large sample, allowing for statistical studies and deriving global conclusions. 

 We find that  LLA GRBs are on average closer than normal lGRBs. Both A$_V$ and N$_H$ of LLAs are similar to those of normal lGRBs. Most LLA GRBs are consistent with the closure relations expected by the fireball model. The few outliers can be accounted for by an early jet break. We show evidences that the events in our LLA sample are also intrinsically fainter during their prompt phase, reinforcing the evidence for a different population. Actually, some events do not follow at all the E$_{p,i}$ - E$_{\rm iso}$ relation.

We have shown that in order to explain all of these properties, we can involve either the geometry of the bursts or a different kind of progenitors. In the former hypothesis, the bursts would be seen slightly off-axis in order to explain the low energy budget observed in these events. In the latter, one could imagine that the progenitor provides less mater for the accretion, thus diminishing the energetic budget at start.

The LLA GRBs sample includes a significant fraction of the supernovae associated with GRBs (64\%), including the best known associations. This means that the conclusion drawn on the general GRB-SN association is based on a sub-sample of the low-luminosity population  (the LLA  GRBs) that might not be representative. We stress the need to confirm this point and the previous work on GRB-SN associations using different spectral and light curve templates, for instance those of GRB 030329/SN2003dh.

\begin{acknowledgements}
We thank Cristiano Guidorzi, at University of Ferrara, for providing help during the analysis of the BAT spectra.
This paper is under the auspices of the FIGARONet collaborative network, supported by the Agence Nationale de la Recherche, program ANR-14-CE33.
We used the data supplied by the UK {\em Swift} Science Data Center at the University of Leicester. 
H. Dereli is supported by the Erasmus Mundus Joint Doctorate Program by Grant Number 2011-1640 from the EACEA of the European Commission. 
\end{acknowledgements}

\

\label{lastpage}

\end{document}